\documentclass[12pt,preprint]{aastex}

\def\cmmd{\hbox{\kern 0.20em cm$^{-2}$}}
\def\htwo{\hbox{H${}_2$}}
\def\cmmt{\hbox{\kern 0.20em cm$^{-3}$}}
\def\msol{\hbox{\kern 0.20em $M_\odot$}}
\def\kms{\hbox{\kern 0.20em km\kern 0.20em s$^{-1}$}}
\def\K{\hbox{\kern 0.20em K}}
\def\mum{\hbox{\kern 0.20em $\mu$m}}
\def\mjy{\hbox{\kern 0.20em mJy}}

\shorttitle{Warm Molecular Hydrogen and ionized Neon in HH~2}
\shortauthors{Lefloch et al.}

\begin{document}

\title{Warm Molecular Hydrogen and Ionized Neon in the HH~2 Outflow
\footnote{Based on observations with ISO, an ESA project with instruments
        funded by ESA Member States (especially the PI countries: France
        Germany, the Netherlands and the United Kingdom) and with the
        participation of ISAS and NASA}
}

\author{B. Lefloch\altaffilmark{1}}
\affil{Laboratoire d'Astrophysique, Observatoire de Grenoble, BP\,53, 
F-38041 Grenoble C\'edex 9, France}
\email{Bertrand.Lefloch@obs.ujf-grenoble.fr}

\author{J.~Cernicharo \altaffilmark{2}}
\affil{Consejo Superior de Investigaciones Cient\'{\i}ficas,
Instituto de Estructura de la Materia, Serrano 123, E-28006 Madrid
}
\author{S.~Cabrit\altaffilmark{3}}
\affil{LERMA, Observatoire de Paris, UMR 8112 du CNRS, France{3}
}
\author{A.~Noriega-Crespo\altaffilmark{4}}
\affil{SIRTF Science Center, Caltech JPL, Pasadena, CA 91125, USA
}
\author{A.~Moro-Mart\'{\i}n\altaffilmark{5}}
\affil{Steward Observatory, University of Arizona, Tucson, AZ, 85721, USA
}
\author{D.~Cesarsky\altaffilmark{6}}
\affil{Max-Planck Institut f\"{u}r Extraterrestrische Physik, 85741 
Garching, Germany
}

\begin{abstract}
We report on spectro-imaging observations of the Herbig-Haro~2 outflow with
the ISOCAM camera onboard the Infrared Space Observatory (ISO). 
The  [\ion{Ne}{2}]~12.81$\mu m$ and  [\ion{Ne}{3}]~15.55$\mu m$ lines are 
detected only
towards the jet working surface (HH~2H), consistent with the high
excitation of this knot in the optical range, while 
\htwo\ pure rotational emission is found all over
the shocked region HH~2. The low energy transition
S(2) traces warm gas {\bf ($\rm T\sim 400\K$)}
peaked towards knots E-F and extended ejecta ($\rm T\sim 250-380\K$) with
masses of a few $10^{-3}\msol$ in the high-velocity CO outflow extending
between the powering source and HH~2. Such emission could arise from 
low-velocity C-type shocks ($\rm v\simeq 10-15\kms$). 
The higher transitions S(3)-S(7)
trace the emission of hot shocked gas ($\rm T= 1000-1400\K$) from
individual optical knots in the HH~2 region.  The ortho to para (OTP) ratio
exhibits large spatial variations between 1.2 (E) and 2.5 (H), 
well below its value at LTE. The emission of the
S(3)-S(7) lines is well accounted for by planar C-shock models with a
typical velocity $\rm V_s= 20-30\kms$ propagating into a medium of density
$\rm n_i= 10^4-10^5\cmmt$ with an initial OTP ratio close to 1 in the 
pre-shock gas. In the
leading edge of the jet, where the geometry of the emission allows a simple
modelling, a good agreement is found with velocities derived from 
the optical proper motions measured in the ionized gas.
\end{abstract}


\section{Introduction}

Bipolar outlows from embedded young stellar objects (YSOs) are perhaps 
one of the most spectacular manifestations of the star formation process. 
One of the best studied and brightest outflows is that of the 
Herbig-Haro (hereafter HH)~1/2 system
(Reipurth 1993). The HH~1/2 system lies in the Orion molecular cloud
at 440 pc and subtends a $\sim 2\arcmin$ angle. The VLA 1 embedded source 
(Pravdo et al. 1985) drives a highly collimated jet that reaches
atomic gas velocities of $\sim 480\kms$~ 
(Eisl\"offel et al. 1994) and produces shocks of $\sim 100\kms$
at its main working surface or {\it bow shock}
(Noriega-Crespo et al. 1989). At optical wavelengths there are at 
least 3 distinct jet flows arising within $5\arcsec$ of the VLA~1 source 
and  the HH~1/2 bow shocks display a complex morphology in high spatial
resolution HST images (Bally et al. 2002). 
The optical jet is associated with a molecular outflow whose high-velocity 
(the "molecular jet") covers deprojected velocities of $15-80\kms$ with 
respect to the ambient cloud (Moro-Mart\'{\i}n et al. 1999). 
Because of these characteristics the HH~1/2 system is a perfect target 
to study  the gas properties in a shock heated environment as well as 
the spatial distribution and the nature of these shocks. 
One of the promises of ISO 
was precisely to be able to discern between the different shocks, either 
C-type or J-type, occurring in a molecular environment. 

We present here spectro-imaging observations of the VLA~1 counter-jet
and the HH~2 region obtained between 5 and $\rm 17\mu m$ with
the ISOCAM camera onboard ISO. We find that the
\htwo\ pure rotational lines S(2) to S(7) arise from 
two physically distinct regions~: a faint extended warm gas component 
associated
with the high-velocity CO jet, seen mainly in S(2), and several more
compact peaks tracing hotter shocked gas in the individual knots A-L of
HH~2 detected in all lines. We report also on the
presence of the [\ion{Ne}{3}]~15.55$\mu m$ and [\ion{Ne}{2}]~12.8$\mu m$ 
lines. HH~2 is the only outflow of low-luminosity 
where these infrared lines have been detected so far
(see Cabrit et al. 1998). 

\section{Observations}
\label{observ}

All observations were obtained with the ISO satellite (Kessler et al.,
1996) and the ISOCAM instrument (Cesarsky et al. 1996). The low resolution
spectra ($\lambda/\Delta\lambda= 40$) between 5 and $\rm 17\mu m$ were
obtained in revolution 691 with the Circular Variable Filter (CVF) with a
pixel scale of $6\arcsec$ and a total field of view of $3\arcmin$ centered
on the HH~2 object. The last pipeline version of the data (OLP10) has been
processed following the package developed at the Institut d'Astrophysique
Spatiale, which removes reasonably well the problem of transients. The
size (HPFW) of the Point Spread Function (PSF) is $\approx 6\arcsec$ for a
pixel scale of $6\arcsec$.

In order to establish accurate astrometry, we used a second CVF map
containing the optically visible Cohen-Schwartz (CS) star, taken in
revolution 873 with a $3\arcsec$ pixel scale.  This second dataset was
processed in the same way as mentioned above.  Unfortunately, internal
reflections between the CVF and the field lens produced spurious ghosts of
the CS star, which is a strong IR emitter, over the full field, preventing
any quantitative analysis of the \htwo\ line emission in the second data
set. However, the presence of the VLA~1 protostar in both data cubes
allows to derive an accurate astrometry for the first CVF image taken with
a $6\arcsec$ pixel scale. The data are presented in Figure 1.
Coordinates are offsets (arcsec) 
relative to the position of VLA1~: $\alpha_{2000}= 05^{h}36^{m}22.6^{s}$, 
$\delta_{2000}= -06^{\circ} 46' 25\arcsec$. 
The interstellar extinction towards
HH~2 was  estimated by Hartmann \& Raymond
(1984), who measured typical reddenings $E(B-V)= 0.11 - 0.44$.  Based on
the extinction curve of Rieke \& Lebofsky (1985), it appears that the flux
dereddening corrections are negligible and we use uncorrected flux values
in what follows.

\section{Results}

\label{results}

Individual CVF spectra towards various optical knots in HH~2 are displayed
in the panels of Fig.~1b. Knot positions are referred by letters A to L
following the nomenclature of Eisloeffel et al. (1994).  At most positions,
the bulk of the emission in the 5-17\mum\ range comes from the \htwo\ pure
rotational lines S(2) to S(7). An exception is the region near
HH~2~H (positions H,B,D) where the fine structure ionic lines 
[\ion{Ne}{2}]~$12.81\mu m$ and [\ion{Ne}{3}] 15.55\mum\ are also detected, 
and even dominate over \htwo\ lines at the nominal position of knot H.  
The spatial distribution of emission flux in the \htwo\ v=0-0 S(2),
S(3), S(5)
lines and in the [Ne II]~12.8\mum\ ionic line is illustrated in Fig.~1a and
Fig.~1c. Three types of morphologies are observed, depending on the line
excitation~:

(i)~The ionic lines [\ion{Ne}{2}] and [\ion{Ne}{3}] show a single peak 
towards knot H (of typical size $\approx 7\arcsec$ at HPFW). 
(ii)~The intermediate excitation lines S(3) and S(5) show two peaks of 
comparable brightness, one
encompassing E, and another peak centered {\sl between} knots H and D, 
shifted by $3\arcsec$ from the ionic line peak. 
(iii)~The low energy S(2) line in Fig.~1a shows a single peak (of
size 13'') encompassing knots E and F. The emission peaks
at the tip of the molecular jet $15\arcsec$ downstream the CO brightness 
peak. The lowest contours of S(2) emission reveal 
a broad pedestal that points towards VLA1/4 (Fig.~1a). Note that the first 
contour is at $5\sigma$ above the map noise level. 
The pedestal overlaps well with the CO ``jet'' 
and suggests that both the CO and \htwo\ emissions are somewhat related, 
although tracing different regions in the ``jet''. 

The overall agreement between the brightness distributions of the pure 
\htwo\ rotational lines and the higher excitation line 
$2.12\mu m$~1-0~S(1) (Davis et al. 1994) is good. Comparison of the line 
intensities indicates that the \htwo\ pure rotational lines are collisionally 
excited (Wolfire \& K\"{o}nigl 1991). 
We show in the right panels of Figure 1 the \htwo\ rotational diagrams 
obtained towards three
representative positions : a region in the CO jet (polygon in Fig.~1a), knot
HH 2E (peak of \htwo\, no ionic line emission) and knot HH 2D (both
ionic and
molecular features).  Overall, the fluxes of the lines $\rm S(J\geq 3)$
match well a linear fit at all positions. The 
excitation temperature $\rm T_{ex}$, the ortho to para ratio (OTP), and the 
total \htwo\ column density of this ``hot'' component are estimated from a 
common linear fit to the temperature for the ortho and para species
in the excitation diagram, and are summarized in Table~1.

Our analysis shows evidence for a hot gas layer with a temperature in the
range 970-1400~K, total \htwo\ column density ranging between 0.18 (H) and
$1.6\times 10^{19}\cmmd$ (E), and OTP ratio ranging between 1.2 (E) and 2.5
(H). There is no spatial trend in the values, i.e. upstream knots do not have
a higher OTP ratio. Interestingly, there appears to be some correlation between
variations in OTP and $\rm T_{ex}$~: the smallest OTP (1.2) is found towards 
HH~2E, which has among
the lowest $\rm T_{ex}$, while the highest OTP (2.5) is found towards HH 2H, 
which has the highest $\rm T_{ex}$. Knots A,C,D, and L fall in between these 
two extremes. Knots B,K,G are the only positions that deviate from this 
trend (low $\rm T_{ex}$ but OTP close to 2).
The OTP ratios measured are systematically lower than the LTE value at 
{\em all} positions but H. It is consistent with the low-excitation 
temperatures measured in the hot gas if \htwo\ is excited by means of a 
C-shock (see e.g. Wilgenbus et al. 2000). 

The rotational diagrams show that, at each position, the population of 
the upper
level of the S(2) line lies well above the hot component fit to the other
transitions. This  implies that the pedestal \htwo\ emission is dominated by 
a second gas component at a lower temperature, discussed in Sect.~4.3

\section{Discussion}
\label{discus}

\subsection{Ionic emission in HH 2H}
The detection of the [\ion{Ne}{2}]12.8\mum\ line towards HH 2H is a clear
signature of J-shocks with velocities above $60\kms$
(Hollenbach \& McKee 1989). The presence of [\ion{Ne}{3}] is consistent
also with the detection of the [\ion{Si}{2}] 34.8\mum~line in HH~2, which 
suggests
shock velocities of $100-140\kms$ (Molinari \& Noriega-Crespo 2002).  That
[\ion{Ne}{2}] and [\ion{Ne}{3}] are detected only towards HH~2H is in line
with the particularly high excitation of this knot in the optical (B\"ohm
\& Solf 1992), and its proper motion larger than $400\kms$
(Bally et al. 2002). The high
excitation of HH~2H is attributed to it being the current location of the
jet working surface (Bally et al. 2002).

The shock speeds inferred in knot H exceed by far the speed at which
molecules are dissociated ($\sim 70\kms$ for a C-shock at $\rm n_0(H)=
10^4\cmmt$; LeBourlot et al. 2002, and $\sim 25\kms$ for a J-shock; 
Hollenbach \& McKee 1989). 
This is consistent with the \htwo\ peak in S(5) being spatially 
shifted from HH~2H.  A similar shift was seen in the \htwo\ 1-0 S(1)
line (Noriega-Crespo \& Garnavich, 1994).  It
indicates that the \htwo\ comes from a separate, lower
velocity shock not physically associated with knot H.

\subsection{Shock-excitation of the hot \htwo\ component in HH knots}

Models of non-dissociative J-shocks (Wilgenbus et al. 2000) or 
dissociative J-shocks with \htwo\ reformation (Flower et al. 2003)
fail to reproduce the CVF data as
the predicted line intensities, expecially S(3)-S(5), are much too weak, or
the required shock velocity is too low ($\sim 10\kms$).  On the contrary,
everywhere in HH~2, the recent models of planar C-shocks 
(Le Bourlot et al. 2002; Cabrit et al. 2003) provide a much better
match with the observations.  

Comparison with the observations favors models with velocity in the 
range $20-30\kms$ and 
a preshock density in the range $10^5-10^4\cmmt$ respectively 
(see the rotational diagrams in Fig.~1). Observations of 
higher-J \htwo\ lines would allow to
better constrain the density in the pre-shock gas. This range of densities 
also agrees with the determinations obtained in the ambient molecular cloud 
(Girart et al.  2002). The OTP ratio in the pre-shock gas is found 
$\simeq 1$. The observed variations in the OTP ratio from knot 
to knot can be produced by changes in the shock speed or preshock 
density, and do not require variations in the initial OTP ratio in the 
preshock gas. 
The characteristic size of the shock emitting region is constrained to be of
the order of $0.8\arcsec-1.7\arcsec$ depending on the individual knots, 
which compares well with their size in the optical. 

The above shock velocities are derived from planal models whereas HST images
provide large evidence for numerous bow-shaped features in HH~2
(Bally et al. 2002). For several knots near the leading edge of HH~2
where the geometry is rather simple and bow-shaped features easily identified
(e.g. F, E, L), the shock velocities compare well with the proper motions 
of the bow wings in the optical (e.g. $50-60\kms$ at E) taking into
account the obliquity of the shock with respect to the direction of the 
motion. 

The absence of spatial trend in the OTP values probably results  from the  
various shocks associated with the mini bows, unresolved in our CVF images. 
Each of these shocks modify the OTP ratio depending on its own excitation 
conditions. 

\subsection{Warm extended $\htwo$ component}

An {\sl upper limit} to the temperature of the extended 
warm component in the \htwo\ pedestal can be derived from the 
ratio of the S(2) and S(3) lines (adopting an OTP ratio 
similar to that in the ambient cloud, close to 1). We find 
$\rm T\leq 420\K$ at the \htwo\ peak and 
$\rm T\leq 380\K$ towards the CO jet. Note that a similar constraint, 
{\em independent of the OTP ratio}, is obtained by assuming that 
at most 50\% of the S(4) flux comes from the pedestal. 

Conversely, in the optically thin limit, we can also
derive a {\sl lower limit} to the temperature by imposing that the
warm component contributes to most of the observed flux of the J= 2-1 
line in the CO "jet" and assuming a standard abundance 
$\rm [CO]/[\htwo]= 10^{-4}$ (Flower and Pineau des For\^ets 1994 showed
indeed that the CO abundance varies little in C-shocks).
This temperature is found by equating expressions of the \htwo\ column
density derived from both tracers. Towards the region of maximum CO
flux ($10-20\K\kms$) in the jet 
the \htwo\ S(2) line flux is $\simeq 8\mjy$/pixel, which yields
a minimum temperature of $\simeq 240-300\K$ in the pedestal.  This
determination depends very little on the adopted OTP value. At 
the \htwo\ peak the CO(2-1) emission is much weaker 
($\simeq 6\K\kms$) and we infer $\rm T \ge 400\K$.

The warm gas at $240-420$~K detected in the pure \htwo\ rotational
lines of low-J is predominantly concentrated towards the tip of the
high-velocity CO jet. The S(2) line suggests a warm gas column 
density of about $2\times 10^{20}\cmmd$ at knots E-F and a corresponding 
mass of $\sim 2\times 10^{-3}\msol$.
The hot gas seen towards the high-velocity CO jet has a 10 times lower
column density (Fig.~1). To reproduce the observed J=4 population, as 
well as the limits on the excitation temperature, a low-velocity 
C-shock with a filling factor of 1, implied by the extended character of 
the S(2) emission, is needed. Then, a rather slow shock
at a velocity of $\sim 10-15\kms$ into gas of density $\rm n(H)\simeq
10^5\cmmt$, with an excitation temperature of  250~K would account for 
the \htwo\ emission of the pedestal along the CO jet. 

Observations of higher~J CO rotational lines at higher angular
resolution are needed to determine more precisely the physical
conditions of the high-velocity CO gas and to compare it with that
traced in S(2) and with molecular shock predictions.

\acknowledgments
J. Cernicharo acknowledges Spanish DGES for this
research under grants PNAYA 2000-1784 and ESP2001-4516.
ANC's research was carried out at the Jet Propulsion Laboratory,
California Institute of Technology, under a contract with NASA;
and partially supported by NASA-APD Grant NRA0001-ADP-096.

\clearpage

\begin{deluxetable}{cccc}
\tablewidth{0pt}
\tablehead{
\colhead{Knot} & \colhead{OTP} & \colhead{$\rm T_h$} & 
\colhead{$\rm N_h(\htwo)$} \\
\colhead{} & \colhead{} & \colhead{(K)} & \colhead{($10^{18}\cmmd$)} \\
}
\tabletypesize{\scriptsize}
\tablecaption{ Parameters of the hot molecular gas derived from 
rotational diagram analysis. 
}
\startdata
A & 1.7 & 1300 & 2.4  \\
B & 1.9 & 1100 & 4.0  \\
C & 1.4 & 1070 & 2.8  \\
D & 1.5 & 1350 & 4.2  \\
E & 1.2 & 1030 & 16  \\ 
G & 2.1 & 1030 & 6.9  \\
H & 2.5 & 1440 & 1.8 \\
K & 2.0 & 970  & 7.2  \\
L & 1.4 & 1070 & 2.7  \\
\enddata
\end{deluxetable}

\clearpage

{\small\noindent Fig.~1 --
{(\em left)}
{\bf (a)~:} Intensity contour map of the \htwo\ 0-0 S(2) line (pink) 
superposed on a [\ion{S}{2}] color image of the HH~1/2 region 
(Reipurth et al. 1993). Contour levels are 4, 7 10, 15, 20, ..., 
$45\mjy/$pixel.
We have superposed the emission of the high-velocity CO outflow in white 
contours. First contour and contour interval are 5 and 2.5~K~km~s$^{-1}$ 
respectively. 
The position of the VLA~1-4 sources is indicated by 
yellow filled circles. 
Black dots mark the location of the knots A-L. 
{\bf (b1-b9)~:} CVF spectra at selected positions. Fluxes are in mJy/pixel. 
The jet spectrum is an average of the individual CV spectra between HH~2 
and VLA~1 in the area of the polygon (yellow) on panel a. The 
wavelength of the pure rotational \htwo\ transitions 
S(2) to S(7) is marked with red ticks.
{\bf (c)~:} Superposed on the same [\ion{S}{2}] picture~: emission contour 
map of the lines $\htwo$ S(3) (c1), $\htwo$ S(5) (c2) and [\ion{Ne}{2}] 
(c3). Contour levels are 0.1, 0.2, ... 0.9 times the brightness peak. 
The \htwo\ intensity contour maps in panels (a) and (c) have been convolved 
with a gaussian of 9\arcsec\ HPFW. \\
{(\em right)}~Rotational diagrams for the \htwo\ rotational states J=3-9. 
The logarithm 
of $\rm N_J/(g_J g_s)$ is plotted against $\rm E_J/k_b$, where $\rm N_J$
is the
beam-averaged column density, $\rm g_J$ is the rotational degeneracy, 
$\rm g_s$ is the spin degeneracy, and $\rm E_J$ is the energy of the 
rotational level J. Open squares mark the data points. Blue straight lines 
show the best fit for a gas component of uniform temperature. 
The gas properties are given in each panel. The dashed lines in each panel
correspond to the same temperature although they differ in the total 
column density for the orto and para species. Red dashed lines trace the fit
to the warm extended component (pedestal).
Green filled stars show the best planar C-shock model that
accounts for the $J\geq 3$ data points. The parameters of the model
(shock velocity $\rm V_s$, hydrogen nuclei density $\rm n_0(H)$ and ortho 
to para 
ratio $\rm otp_0$ in the pre-shock gas) and the pixel filling factor ff
are given for HH~2E and HH~2D.
Below the panels of HH~2E and HH~2D  are given the best fit parameters 
for planar C-shock models.
}
\clearpage

\begin{figure}
\epsscale{1.0}
\caption{}
\label{fig1}
\end{figure}
\clearpage

\end{document}